\input harvmac
%
% S-Tables Macro
%
\message{S-Tables Macro v1.0, ACS, TAMU (RANHELP@VENUS.TAMU.EDU)}
%
% Help Text
%
\newhelp\stablestylehelp{You must choose a style between 0 and 3.}%
\newhelp\stablelinehelp{You
should not use special hrules when stretching
a table.}%
\newhelp\stablesmultiplehelp{You have tried to place an S-Table
inside another
S-Table.  I would recommend not going on.}%
%
% Line Thicknesses (Values)
%
\newdimen\stablesthinline
\stablesthinline=0.4pt
\newdimen\stablesthickline
\stablesthickline=1pt
%
% Border and Internal Line Thicknesses
%
\newif\ifstablesborderthin
\stablesborderthinfalse
\newif\ifstablesinternalthin
\stablesinternalthintrue
\newif\ifstablesomit
\newif\ifstablemode
\newif\ifstablesright
\stablesrightfalse
%
% Save Registers
%
\newdimen\stablesbaselineskip
\newdimen\stableslineskip
\newdimen\stableslineskiplimit
%
% Counters
%
\newcount\stablesmode
\newcount\stableslines
\newcount\stablestemp
\stablestemp=3
\newcount\stablescount
\stablescount=0
\newcount\stableslinet
\stableslinet=0
%
% Table Style Selection
%
% 0 - Centered
% 1 - Left Justified
% 2 - Right Justified
% 3 - Not Justified
%
\newcount\stablestyle
\stablestyle=0
%
% Element Buffering Definitions
%
\def\stablesleft{\quad\hfil}%
\def\stablesright{\hfil\quad}%
%
% Vertical Bar Activation
%
\catcode`\|=\active%
%
% Strut Control
%
\newcount\stablestrutsize
\newbox\stablestrutbox
\setbox\stablestrutbox=\hbox{\vrule height10pt depth5pt width0pt}
\def\stablestrut{\relax\ifmmode%
                         \copy\stablestrutbox%
                       \else%
                         \unhcopy\stablestrutbox%
                       \fi}%
%
% Misc. Internal Stuff
%
\newdimen\stablesborderwidth
\newdimen\stablesinternalwidth
\newdimen\stablesdummy
\newcount\stablesdummyc
\newif\ifstablesin
\stablesinfalse
%
% Table Macros
%
\def\begintable{\stablestart%
  \stablemodetrue%
  \stablesadj%
  \halign%
  \stablesdef}%
\def\stablesadj{%
  \ifcase\stablestyle%
    \hbox to \hsize\bgroup\hss\vbox\bgroup%
  \or%
    \hbox to \hsize\bgroup\vbox\bgroup%
  \or%
    \hbox to \hsize\bgroup\hss\vbox\bgroup%
  \or%
    \hbox\bgroup\vbox\bgroup%
  \else%
    \errhelp=\stablestylehelp%
    \errmessage{Invalid style selected, using default}%
    \hbox to \hsize\bgroup\hss\vbox\bgroup%
  \fi}%
\def\stablesend{\egroup%
  \ifcase\stablestyle%
    \hss\egroup%
  \or%
    \hss\egroup%
  \or%
    \egroup%
  \or%
    \egroup%
  \else%
    \hss\egroup%
  \fi}%
\def\stablestart{%
  \ifstablesin%
    \errhelp=\stablesmultiplehelp%
    \errmessage{An S-Table cannot be placed within an S-Table!}%
  \fi
  \global\stablesintrue%
  \global\advance\stablescount by 1%
  \message{<S-Tables Generating Table \number\stablescount}%
  \begingroup%
  \stablestrutsize=\ht\stablestrutbox%
  \advance\stablestrutsize by \dp\stablestrutbox%
  \ifstablesborderthin%
    \stablesborderwidth=\stablesthinline%
  \else%
    \stablesborderwidth=\stablesthickline%
  \fi%
  \ifstablesinternalthin%
    \stablesinternalwidth=\stablesthinline%
  \else%
    \stablesinternalwidth=\stablesthickline%
  \fi%
  \tabskip=0pt%
  \stablesbaselineskip=\baselineskip%
  \stableslineskip=\lineskip%
  \stableslineskiplimit=\lineskiplimit%
  \offinterlineskip%
  \def\borderrule{\vrule width \stablesborderwidth}%
  \def\internalrule{\vrule width \stablesinternalwidth}%
  \def\thinline{\noalign{\hrule height \stablesthinline}}%
  \def\thickline{\noalign{\hrule height \stablesthickline}}%
  \def\trule{\omit\leaders\hrule height \stablesthinline\hfill}%
  \def\ttrule{\omit\leaders\hrule height \stablesthickline\hfill}%
  \def\tttrule##1{\omit\leaders\hrule height ##1\hfill}%
  \def\stablesel{&\omit\global\stablesmode=0%
    \global\advance\stableslines by 1\borderrule\hfil\cr}%
  \def\el{\stablesel&}%
  \def\elt{\stablesel\thinline&}%
  \def\eltt{\stablesel\thickline&}%
  \def\elttt##1{\stablesel\noalign{\hrule height ##1}&}%
  \def\elspec{&\omit\hfil\borderrule\cr\omit\borderrule&%
              \ifstablemode%
              \else%
                \errhelp=\stablelinehelp%
                \errmessage{Special ruling will not display properly}%
              \fi}%
  \def\stmultispan##1{\mscount=##1 \loop\ifnum\mscount>3
\stspan\repeat}%
  \def\stspan{\span\omit \advance\mscount by -1}%
  \def\multicolumn##1{\omit\multiply\stablestemp by ##1%
     \stmultispan{\stablestemp}%
     \advance\stablesmode by ##1%
     \advance\stablesmode by -1%
     \stablestemp=3}%
  \def\multirow##1{\stablesdummyc=##1\parindent=0pt\setbox0\hbox\bgroup%

    \aftergroup\emultirow\let\temp=}
  \def\emultirow{\setbox1\vbox to\stablesdummyc\stablestrutsize%
    {\hsize\wd0\vfil\box0\vfil}%
    \ht1=\ht\stablestrutbox%
    \dp1=\dp\stablestrutbox%
    \box1}%

\def\stpar##1{\vtop\bgroup\hsize ##1%
     \baselineskip=\stablesbaselineskip%
     \lineskip=\stableslineskip%

\lineskiplimit=\stableslineskiplimit\bgroup\aftergroup\estpar\let\temp=}%

  \def\estpar{\vskip 6pt\egroup}%
  \def\stparrow##1##2{\stablesdummy=##2%
     \setbox0=\vtop to ##1\stablestrutsize\bgroup%
     \hsize\stablesdummy%
     \baselineskip=\stablesbaselineskip%
     \lineskip=\stableslineskip%
     \lineskiplimit=\stableslineskiplimit%
     \bgroup\vfil\aftergroup\estparrow%
     \let\temp=}%
  \def\estparrow{\vfil\egroup%
     \ht0=\ht\stablestrutbox%
     \dp0=\dp\stablestrutbox%
     \wd0=\stablesdummy%
     \box0}%
  \def|{\global\advance\stablesmode by 1&&&}%
  \def\|{\global\advance\stablesmode by 1&\omit\vrule width 0pt%
         \hfil&&}%
  \def\vt{\global\advance\stablesmode by 1&\omit\vrule width
\stablesthinline%
          \hfil&&}%
  \def\vtt{\global\advance\stablesmode by 1&\omit\vrule width
\stablesthickline%
          \hfil&&}%
  \def\vttt##1{\global\advance\stablesmode by 1&\omit\vrule width ##1%
          \hfil&&}%
  \def\vtr{\global\advance\stablesmode by 1&\omit\hfil\vrule width%
           \stablesthinline&&}%
  \def\vttr{\global\advance\stablesmode by 1&\omit\hfil\vrule width%
            \stablesthickline&&}%
  \def\vtttr##1{\global\advance\stablesmode by 1&\omit\hfil\vrule
width ##1&&}%
  \stableslines=0%
  \stablesomitfalse}
\def\stablesdef{\bgroup\stablestrut\borderrule##\tabskip=0pt plus 1fil%
  &\stablesleft##\stablesright%
  &##\ifstablesright\hfill\fi\internalrule\ifstablesright\else\hfill\fi%

  \tabskip 0pt&&##\hfil\tabskip=0pt plus 1fil%
  &\stablesleft##\stablesright%
  &##\ifstablesright\hfill\fi\internalrule\ifstablesright\else\hfill\fi%

  \tabskip=0pt\cr%
  \ifstablesborderthin%
    \thinline%
  \else%
    \thickline%
  \fi&%
}%
\def\endtable{\advance\stableslines by 1\advance\stablesmode by 1%
   \message{- Rows: \number\stableslines, Columns:
\number\stablesmode>}%
   \stablesel%
   \ifstablesborderthin%
     \thinline%
   \else%
     \thickline%
   \fi%
   \egroup\stablesend%
\endgroup%
\global\stablesinfalse}
%
% end of STABLES.TEX

% -------------------------AK Definitions
\font\cmss=cmss10 \font\cmsss=cmss10 at 7pt
 \def\inbar{\,\vrule height1.5ex width.4pt depth0pt}
\def\IZ{\relax\ifmmode\mathchoice
{\hbox{\cmss Z\kern-.4em Z}}{\hbox{\cmss Z\kern-.4em Z}}
{\lower.9pt\hbox{\cmsss Z\kern-.4em Z}}
{\lower1.2pt\hbox{\cmsss Z\kern-.4em Z}}\else{\cmss Z\kern-.4em
Z}\fi}
\def\IB{\relax{\rm I\kern-.18em B}}
\def\IC{{\relax\hbox{$\inbar\kern-.3em{\rm C}$}}}
\def\ID{\relax{\rm I\kern-.18em D}}
\def\IE{\relax{\rm I\kern-.18em E}}
\def\IF{\relax{\rm I\kern-.18em F}}
\def\IG{\relax\hbox{$\inbar\kern-.3em{\rm G}$}}
\def\IGa{\relax\hbox{${\rm I}\kern-.18em\Gamma$}}
\def\IH{\relax{\rm I\kern-.18em H}}
\def\II{\relax{\rm I\kern-.18em I}}
\def\IK{\relax{\rm I\kern-.18em K}}
\def\IP{\relax{\rm I\kern-.18em P}}
\def\IR{\relax{\rm I\kern-.18em R}}

%----------------------------

\lref\lsmt{W. Lerche, C. Schweigert, R. Minasian, and
S. Theisen, hep-th/9711104.}
\lref\bianchi{M. Bianchi, hep-th/9711201.}
\lref\dh{A. Dabholkar and J. Harvey, \lq\lq Nonrenormalization of
the Superstring Tension,'' Phys. Rev. Lett. {\bf 63} (1989) 478.}
\lref\sw{N. Seiberg and E. Witten, hep-th/9603003.}
\lref\gh{O. Ganor and A. Hanany, hep-th/9602120.}
\lref\CHL{S. Chaudhuri, G. Hockney, and J. Lykken, Phys. Rev. Lett.
{\bf 75} (1995) 2264.}
\lref\cp{
S. Chaudhuri and J. Polchinski, Phys. Rev. {\bf D52} (1995) 7168.}
\lref\DP{A. Dabholkar and J. Park, Nucl. Phys. {\bf B477} (1996) 701.}
\lref\kv{S.\ Kachru and C.\ Vafa, Nucl. Phys. {\bf B450} (1995) 69,
hep-th/9505105.}
\lref\klm{
A. Klemm, W. Lerche and P. Mayr, Phys. Lett. {\bf B357} (1995) 313\semi
P. S. Aspinwall and J. Louis, Phys. Lett. {\bf B369} (1995) 233.}
\lref\aspgr{P. Aspinwall and M. Gross, hep-th/9605131.}
\lref\bikmsv{ M. Bershadsky, K. Intriligator, S. Kachru, D. Morrison,
V. Sadov, and C. Vafa, \lq\lq Geometric Singularities and Enhanced
Gauge Symmetries,'' hep-th/9605200.}
\lref\vafawit{C. Vafa and E. Witten, hep-th/9507050.}

\lref\lg{
C.\ Vafa, Mod. Phys. Lett. {\bf A4} (1989) 1169;
K.\ Intriligator and C.\ Vafa,  \np339 (1990) 95 }
\lref\kmv{A.\ Klemm, P.\ Mayr and C.\ Vafa, Nucl. Phys. B,
Proc. Suppl. 58 (1997) 177-194, hep-th/9607139}
\lref\fhpv{P.\ Forgacs, Z.\ Horv\`ath, L.\ Palla and P.\ Vecserny\'es,
    \np308(1988)477}
\lref\hktyIaII{S.\ Hosono, A.\ Klemm, S.\ Theisen and S.\ T.\ Yau,
      \cmp167(1995)301, hep-th/9308122 and
      \np433(1995)501, hep-th/9406055}
\lref\batyrev{V. Batyrev, Duke Math. Journ {\bf 69} (1993) 349}

\lref\reidI{M.\ Reid, {\sl Canonical 3-folds}, { Journe\'es de
                G\'eometrie Alg\'ebrique d' Angers}, Sijthoff \&
Nordhoff,
                 1980, 273}
\lref\reidII{ M.\ Reid, {\sl Young Person's guide to canonical
          Singularities}, in {\sl Algebraic Geometry: Bowdoin},
          Proc. of Symp. in Pure Math. {\bf 46} (AMS, 1987)}
\lref\bs{V. Batyrev and  D. van Straten, \cmp 168(1995)493
alg-geom/9307010}

\lref \cf{P.\ Candelas and A.\ Font, hep-th/9603170, E.\ Perevalov and 
H. \ Skarke, hep-th/9704129}
\lref\grschw{M. Green, J. Schwarz, and P. West, Nucl. Phys.
{\bf B254} (1985) 327.}

\Title{\vbox{\baselineskip12pt\hbox{hep-th/9712035}
\hbox{EFI-97-55, LBNL-41125, UCB-PTH-97/59}
}}
{\vbox{\centerline{
Calabi-Yau Duals for CHL Strings}}
}
\centerline{Shamit Kachru$^{1}$, Albrecht Klemm$^{2}$,
and Yaron Oz$^{1}$}
\bigskip
\bigskip
\centerline{$^{1}$Department of Physics}
\centerline{University of California at Berkeley}
\smallskip
\centerline{and}
\smallskip
\centerline{Theoretical Physics Group}
\centerline{Ernest Orlando Lawrence Berkeley National Laboratory}
\centerline{University of California}
\centerline{Berkeley, California 94720}
\medskip
\medskip
\centerline{$^{2}$ Enrico Fermi Institute}
\centerline{University of Chicago}
\centerline{5460 S. Ellis Avenue}
\centerline{Chicago, IL 60637}
\bigskip
\bigskip
\noindent
We find M-theory (Type IIA) duals for
compactifications
of the 9d CHL string to 5d (4d) on $K3$ ($K3 \times S^1$).
The IIA duals are Calabi-Yau orbifolds with nontrivial
RR $U(1)$ backgrounds turned on.

\Date{November 1997}
%\draftmode

\newsec{Introduction}

In ten dimensions, the consistent critical
string theories with (at least) sixteen supercharges
have been known since the 1980s.
There are (after accounting for the S-duality of the
two $SO(32)$ theories) four.
In $D=9$, in addition to the compactifications of the $D=10$
theories on $S^1$ we find a new theory with 16 supercharges --
the CHL string \refs{\CHL,\cp}.  This theory can be
obtained by compactifying the $E_8 \times E_8$ heterotic
string on a circle with radius $R^9$ and
orbifolding by the $Z_2$ which identifies the two $E_8$s and
shifts $x^9 \rightarrow x^9 + \pi R^9$.
The resulting heterotic string has $E_8$ current algebra at level
2 on its worldsheet.  The moduli space of vacua
in nine-dimensions is
\eqn\modspace{SO(9,1;\IZ)\backslash SO(9,1) / SO(9)\times SO(1)~.}
The low-energy M-theoretic description of this theory involves
the compactification of 11 dimensional supergravity on a
M\"obius strip \DP.

\lref\strstr{C. Hull and P. Townsend, Nucl. Phys. {\bf B438} (1995)
109\semi E. Witten, Nucl. Phys. {\bf B443} (1995) 85.}
\lref\fhsv{S. Ferrara,
J. Harvey, A. Strominger, and C. Vafa, Phys. Lett. {\bf B361} (1995)
59,  hep-th/9505162.}
\lref\mv{
D. Morrison and C. Vafa, hep-th/9602114, hep-th/9603161.}
\lref\bikmsv{M. Bershadsky, K. Intriligator, S. Kachru, D. Morrison,
V. Sadov, and C. Vafa, hep-th/9605200. }
\lref\chldual{J. Schwarz and A. Sen, Phys. Lett. {\bf B357} (1995) 323,
hep-th/9507027.}

In the case of the $E_8 \times E_8$ string,
dual Type IIA descriptions after compactification to 6d
on a torus or 4d on
$K3 \times T^2$ were given in \refs{\strstr,\kv,\fhsv,\mv}.
For the CHL string, the dual of the maximally supersymmetric
compactification to 6d was given by Schwarz and Sen \chldual.
Other aspects of the maximally supersymmetric CHL compactifications
were recently discussed in \refs{\lsmt,\bianchi}.
It is the purpose of this paper to begin the task of finding
duals of CHL compactifications with less
supersymmetry, by finding the IIA and M-theory duals of the CHL
compactifications to 4 and 5 dimensions with 8 supercharges.

\lref\kkotwo{S. Kachru, A. Klemm, and Y. Oz, to appear.}
\lref\web{B. Greene, D. Morrison, and A. Strominger,
Nucl. Phys. {\bf B451} (1995) 109, hep-th/9504145\semi
T. Chiang, B. Greene, M. Gross, and Y. Kanter, hep-th/9511204\semi
A. Avram, P. Candelas, D. Jancic, and M. Mandelberg,
Nucl. Phys. {\bf B465} (1996) 458, hep-th/9511230.}
In investigations of
vacua of the
$E_8 \times E_8$ heterotic string with 8 supercharges, a proper
understanding of $\it singular$ points in the moduli space
has led to the discovery of many new nontrivial renormalization
group
fixed points in $d=4,5,6$ (in \refs{\gh,\sw} and much subsequent work).
Analysis of the dual Calabi-Yau models has been a powerful tool
for exploring these field theories.  One motivation for our
work is to provide a similar framework for studying novel theories 
without gravity which may arise in $d=4,5$ from CHL compactifications.
Results in
this direction will appear in a companion paper \kkotwo.
It would also be interesting to determine whether or not the
web of 4d $N=2$ string vacua discussed here
is connected, through phase transitions, to the web of
conventional 4d $N=2$ Calabi-Yau
vacua \web.

\newsec{The Heterotic Theories}

Starting from the 9d CHL string with $E_8 \times U(1)^2$ gauge symmetry,
one can compactify on $K3\times S^1$ and obtain a 4d $N=2$
supersymmetric
low-energy theory.  From the perturbative heterotic string Bianchi
identity,
one should in addition require a background gauge bundle
$V$ on the $K3$ with
\eqn\chern{c_{2}(V) = 12~.}
Said differently, there should be 12 instantons embedded in the $E_8$.

One argument that \chern\ is the correct condition is the following.
Start with the $E_8 \times E_8$ string on $K3 \times T^2$.
Imagine there are bundles $V_{1,2}$ embedded in the two $E_8$s.
Then, one can do a $Z_2$ orbifold to obtain the CHL string as long as
$V_1$ and $V_2$ are identical.
The Bianchi identity for the $E_8 \times E_8$ theory in this
case is
\eqn\bian{c_{2}(V_{1}) + c_{2}(V_2) = 24}
so in particular to do the CHL orbifold, one requires $c_{2}(V_{1,2}) =
12$.  After orbifolding, one is left with the diagonal $E_8$ and
a bundle of instanton number 12.

\lref\dmw{M. Duff, R. Minasian, and E. Witten,
Nucl. Phys. {\bf B465} (1996) 413, hep-th/9601036.}
A more microscopic description of the same vacuum comes from M-theory on
a M\"obius strip $M$ times $K3\times S^1$.  In
this description, the $E_8$ gauge
fields and the bundle $V$ live on $\partial M$.  One can now imagine
more general configurations where $N$ instantons shrink and leave
the boundary of the world in their avatar as fivebranes wrapping
the base of $M$ and the circle.\foot{By the base of $M$ we
mean a representative of the nontrivial class in $H_{1}(M)$, which
looks like a base if one locally views $M$ as a fibration
of the interval over $S^1$.}
Then,
the Bianchi identity \chern\ is modified and
becomes
\eqn\modchern{c_{2}(V) + N = 12~.}
This can be argued as in the previous paragraph, now by using
the general configurations studied by Duff, Minasian, and
Witten \dmw.
We will mostly concentrate on finding duals in the
case that $c_{2}(V) = 12$,
though we also find Calabi-Yau duals
for some cases with wrapped fivebranes
present (notably, the maximal case $N=12$).

For a fixed choice of instanton and fivebrane numbers satisying
\modchern,
we can
still find a whole web of vacua by considering bundles with
different structure groups, yielding different unbroken
non-Abelian gauge groups $G \subset E_8$.
By passing to a generic point on
the Coulomb branch of $G$,
one obtains an Abelian gauge theory characterized by the number
of vector and hypermultiplets.  Some of the expected gauge groups
and matter contents are presented below in Table 1, in terms of
\eqn\defn{n ~\equiv ~8-N ~= ~c_{2}(V) - 4~.}
One should consider
\eqn\ineq{n \geq 0~}
in the table, since for $n<0$ there are no stable bundles with
the right instanton number on $K3$.  The number of hypermultiplets
$n_H$ at a generic point in the Coulomb branch of $G$ is given below,
while the number of vector multiplets at a generic point is given
by
\eqn\numvec{n_V = {\rm rank(G)} + N + 3~.}
It is important to emphasize that we only list the unbroken
subgroups of the perturbative $E_8$ in Table 1;  the table
ignores the omnipresent $U(1)^3$ which appears in \numvec,
as well as the $N$ vector multiplets on the (wrapped) fivebrane
worldvolumes.

In addition to listing the gauge group $G$, we have presented
the relevant singularity type expected to produce $G$ in the
Calabi-Yau dual, and the expected $G$ charged matter content
(which becomes massless at the origin of the $G$ Coulomb branch,
in the $\it classical$ heterotic string theory).
The charged matter content is simply computed using $c_{2}(V)$
and the index theorem, as in \kv.
Decompose the adjoint of $E_8$ under $G\times H$ (where
$H$ is the commutant of $G$ in $E_8$) as
\eqn\adjdec{{\bf 248} = \sum_i (M_i, R_i)}
where $M_i$ is the $G$ representation and $R_i$ is the
$H$ representation. Then it follows from the index theorem
that the number of left-handed spinor multiplets transforming
in the $M_i$ representation of $G$ is
given by
\eqn\index{N_{M_i} ~=~{\rm dim(R_i)} - {1\over 2} \int_{K3} c_{2}(V)
~{\rm index(R_i)}~.}
\index\ is normalized to properly count numbers of hypermultiplets.

The detailed explanation
of the singularity types (and in particular the occurence of
\lq\lq split," \lq\lq non-split," and \lq\lq semi-split"
singularities, denoted with superscript $s$, $ns$, and $ss$) can
be found in \bikmsv, from which Table 1 has been lifted with
suitable modifications.  The geometrical realization of
non-simply laced gauge groups was first explained in \aspgr.
Roughly speaking, the threefold singularities can be understood
as elliptic surface singularities fibered over an additional curve.
There can be monodromies which orbifold the naive (A-D-E)
gauge group coming
from the surface singularity by an outer automorphism as one goes
around singular points on the additional curve, yielding a
non-simply laced group.  In the \lq\lq split" cases this does
$\it not$ occur, while in the other cases such an outer automorphism
$\it does$ act.

In all cases in Table 1 where a multiplicity becomes
negative, the corresponding branch with gauge group $G$ does not
exist (there aren't enough instantons to break $E_8$ to $G$).
In addition, the complete Higgsing of $E_8$ is only possible for
$n \geq 6$.

Table 1: Some Strata in the Moduli Space
\bigskip
\begintable
 Type | Group | Matter content | $n_H$ \elt $E_7$ | $E_7$ |
 $({n \over 2}){\rm \bf 56}$ | $n+33$ \elt $E_6^{s}$ | $E_6$
 |$(n-2){\rm \bf 27}$ | $2n+32$ \elt $E_6^{ns}$ | $F_4$ | $(n-3){\rm
 \bf 26}$| $3n+30$ \elt $D_5^{s}$ | $SO(10)$ | $(n-4){\rm \bf
 16}+(n-2){\rm \bf 10}$ | $3n+29$ \elt $D_5^{ns}$ | $SO(9)$ |
 $(n-3){\rm \bf 9}+(n-4){\rm \bf 16}$ | $5n+24$ \elt $D_4^{s}$ |
 $SO(8)$ | $(n-4)({\rm \bf 8}_c+{\rm \bf 8}_s+{\rm \bf 8}_v)$ |
 $5n+24$
\elt
 $D_4^{ss}$ | $SO(7)$ | $(n-5){\rm \bf 7}+(2n-8){\rm \bf 8}$ | $7n+15$
 \elt $D_4^{ns}$ | $G_2$ | $(3n-14){\rm \bf 7}$ | $11n-2$ \elt
$A_3^{s}$ |
 $SU(4)$ | $(n-6){\rm \bf 6}+(4n-16){\rm \bf 4}$ | $7n+15$ \elt
$A_3^{ns}$
 | $SO(5)$ | $(n-7){\rm \bf 5}+(4n-16){\rm \bf 4}$ | $9n+2$ \elt
$A_1\times A_1$ | $SO(4)$ | $(n-8) {\rm \bf (2,2)} + (4n-16) [{\rm \bf
(1,2) +
(2,1)}]$ | $9n + 2$ \elt
$A_2^{s}$ | $SU(3)$ | $(6n-30){\rm \bf 3}$ | $11n-2 $ \elt $A_1$ |
 $SU(2)$ | $(6n-32){\rm \bf 2}$ | $17n-33$\elt
 $D_6^{s}$ | $SO(12)$   | ${r \over 2}{\rm \bf 32}+
                        ({n-r-4 \over 2}){\rm \bf 32}^{\prime}+
                             (n){\rm \bf 12}$  | $n+30$ \elt
 $D_6^{ns}$ | $SO(11)$  | $({n \over 2}-2){\rm \bf 32}+
                      (n-1){\rm \bf 11}$  | $3n+29$ \elt
 $A_5^{s}$ | $SU(6)$   | ${r \over 2} {\bf 20}+
     (r+2n) {\bf 6}+
           (n-r-6){\bf  15}$ |  $2n-r+25$ \elt
 $A_5^{ns}$  |  $Sp(3)$      |  $(2n+{3\over 2}r){\bf 6}
+(n-r-7){\bf 14}
+\half r {\bf 14'}$   | $3n+19-2r$ \elt
 $A_4^{s}$ | $SU(5)$   | $(3n-8){\rm \bf 5}+(n-6){\rm \bf 10}$ |
        $4n+24$ \elt
${\rm None}$ | ${\rm None}$ |$(~only~ possible~ for~ n \geq 6~)$|$29n-
100$

\endtable

\bigskip
In the case $G=SO(12)$,
an
additional integer $r$ is required to
specify the heterotic vacuum.  That is because the
commutant of $SO(12)$ in $E_8$ is $SO(4) \simeq SU(2)\times SU(2)$,
and we can embed $r+4$ instantons in one $SU(2)$ and
$n-r$ instantons in the other $SU(2)$.
In order to keep at least 4 instantons in each $SU(2)$ (which
is the minimum number for which a suitable $SU(2)$ bundle exists),
we must require $r \geq 0$ and $(n-r) \geq 0$.
The cases $G=SU(6), Sp(3)$ are also $r$-dependent, because
they can be obtained by Higgsing $SO(12)$ for suitable $r$, but
not by Higgsing $E_7$ with ${\bf 56}$s.

\newsec{Calabi-Yau Duals}
\subsec{Method for Finding Calabi-Yau Duals}

There is a well known string-string duality relating the
heterotic string on $T^4$ to type IIA compactifications on
$K3$.  Schwarz and Sen found an analogous statement for the
compactification of the 9d CHL string to six dimensions on a
$T^3$ \chldual.
Namely, they found that this 6d CHL string is dual to
Type IIA on a $Z_2$ orbifold of $K3$ which preserves precisely
$\it twelve$ of the twenty $(1,1)$ forms.  The other eight are
projected out of the untwisted sector.  This exactly reproduces
the rank reduction (by eight) of the CHL string.

Normally, the 8 \lq\lq missing" $(1,1)$ forms would be resurrected
in the twisted sector.  However, by also embedding a $Z_2$ action in
a RR $U(1)$ gauge group, they were able to remove the twisted sector
contributions.  The $Z_2$ gauge flux is concentrated at the
orbifold fixed points
of the geometrical $Z_2$ action on the $K3$, and removes the blow-up
modes.
The conclusion is that IIA on this particular $K3/Z_2$ is dual to
the CHL string.

As was discussed in \chldual, a more geometrical formulation of the
same $Z_2$ action can be given in M-theory.  The $E_8 \times E_8$
heterotic string on
$T^4$ is dual to M-theory on $K3 \times S^1$.  If we go to the
$E_8 \times E_8$ point in moduli space and do the CHL orbifold,
the exchange of the two $E_8$s maps in M-theory to a $Z_2$ which
exchanges the two $E_8$ singularities of the $K3$.
If we take the $K3$ to be an elliptic fibration with coordinate
$z$ on the $P^1$ base and a defining equation of
the form
\eqn\kthreed{y^{2} = x^{3} + xf(z) + g(z)}
then the $Z_2$ acts by
\eqn\ztwo{z \rightarrow {1\over z}, ~y \rightarrow -y~.}
The two $E_8$ singularities naturally arise at $z = 0, \infty$ and
are identified by \ztwo.
The shift on the heterotic
circle maps, in M-theory, to a shift on the M-theory $S^1$.  Therefore,
the $Z_2$ symmetry is $\it freely~acting$ in M-theory, and the
possibility
of fixed point contributions to the spectrum does not arise.
In the IIA picture, the $U(1)$ RR gauge field arises via Kaluza-Klein
reduction along the M-theory circle, and the shift on the $S^1$
therefore
embeds a $Z_2$ action in the $U(1)$.

We are interested in using a similar technique to find Calabi-Yau duals
for 9d CHL strings compactified to 4d on $K3 \times S^1$, or to 5d on
$K3$.  We can use the above duality and the adiabatic argument
\vafawit\ to
find duals as follows.

Consider the heterotic $E_8 \times E_8$ string
on $K3 \times T^2$ with (12,12) instantons
in the two $E_8$s.
After maximal Higgsing, it is known to be dual to the
IIA compactification on the Calabi-Yau $X$ given by an elliptic
fibration over $P^1 \times P^1$ with hodge numbers $(3,243)$.
If we Higgs \lq\lq symmetrically," then we can still perform the
CHL $Z_2$ by exchanging the two $E_8$s and shifting one of the
circles of the $T^2$.  This will leave us with a CHL string
compactification
with 132 hypermultiplets and 3 vector multiplets.  The corresponding
Calabi-Yau dual $X_{CHL}$ should have hodge numbers $(3,131)$.
More accurately, these should be the contributions to the hodge numbers
from the \lq\lq untwisted" cohomology classes, ignoring any
modes which originate at the $Z_2$ fixed loci on $X_{CHL}$.

Using the adiabatic argument, we can construct $X_{CHL}$ from $X$.
$X$ is a $K3$ fibration, and we can implement the action
\ztwo\ on the $K3$ fibers of $X$, while at the same time acting on
the $P^1$ in a way that preserves 4d $N=2$ supersymmetry
(the precise details are provided in \S3.2).
If we consider M-theory on $X \times S^1$ and at the same time
act with a shift on the $S^1$, then the overall $Z_2$ action will
be free.  This is why the massless modes should
come from the \lq\lq untwisted
sector'' of $X_{CHL} = X/Z_2$.  In fact, one finds agreement with
the expected $(3,131)$.

\lref\edphases{E. Witten, Nucl. Phys. {\bf B471}
(1996) 195, hep-th/9603150.}
Given that M-theory on $X_{CHL} \times S^1$ is dual to the 4d
$N=2$ CHL compactification, we can now make the circle very large
and obtain an approximate 5d duality between M-theory on $X_{CHL}$
(with the prescription that the $Z_2$ fixed points cannot be resolved)
and the compactification of the 9d CHL string on $K3$.
More precisely, we need to take a double-scaling limit to go to
the M-theory description, as in \edphases.  The Kahler classes of
$X_{CHL}$ as measured in Type IIA and M-theory, which we
will call $K_{IIA}$ and $K_{M}$,  are related by
\eqn\edeq{K_M = {1\over {T^{1/3}R}} K_{IIA}}
where $T$ is the two-brane tension and $R$ is the radius of the
circle.  If we take $R$ to
infinity but wish to keep the Kahler classes of
the Calabi-Yau fixed in M-theory units, we must take $K_{IIA}
\rightarrow
\infty$ as well.
Note that although $X$ also admits an \lq\lq F-theory limit," which is
dual to an $E_8\times E_8$ compactification on $K3$, $X_{CHL}$ does not.
The F-theory limit involves shrinking the elliptic fibers of $X$, and
the $Z_2$ action which turns $X$ into $X_{CHL}$ destroys the relevant
fibration structure.
This is not surprising, since there is no obvious way to get theories
with 6d $(1,0)$ supersymmetry starting with the 9d CHL string.

Alternatively, as in \chldual,
we can view the $S^1$ shift as a $Z_2$ RR flux in IIA string theory.
Then we see that IIA on $X_{CHL}$, with suitable $Z_2$ fluxes at the
orbifold fixed points, is dual to the maximally Higgsed
4d $N=2$ CHL vacuum.

More generally, one can first \lq\lq unHiggs" some gauge group $G$
by going through suitable extremal transitions starting with $X$.  Of
course, one should unHiggs symmetrically in the two $E_8$s.
Then, using the same fibre-wise argument to find the correct
$Z_2$ action, we can find Calabi-Yau duals for the general CHL
compactifications on $K3$ discussed in \S2.
We have done this in many cases, and find complete agreement with
the CHL expectations.

Generally, $N=2$ 
supergravity in 4d requires a local integrability structure on 
the vector moduli space known as special geometry (globally,
it requires the positivity 
of the kinetic terms). Both 
requirements are naturally satisfied for the moduli 
spaces of Calabi-Yau threefolds.  A natural generalisation is to consider 
the moduli space of an invariant sector under a group action on the 
Calabi-Yau.  This is familiar 
for rigid $N=2$ theories, where one has to consider quite
generally certain subsectors of the moduli space of a Riemann surface 
and not the full abelian variety, but a Prym variety. However, 
unlike the 
Riemann surfaces,
the CY is not an auxiliary surface for decribing the moduli 
space. 
Hence, in general twisted sectors have to be considered, 
and one ends up with the
moduli space of another CY. The CHL string gives a rationale for 
dealing only with invariant states 
under particular $Z_2$ actions. It would be very
interesting to see whether other invariant sectors 
of Calabi-Yau moduli spaces 
can also arise in special string constructions.

\subsec{Toric Description of the Calabi-Yau Duals}

As described in \S3.1, to construct a dual description of the
$N=2$ CHL vacua in four dimensions, we start with the 
non-perturbative equivalence between the $E_8 \times E_8$ heterotic
string on $K3 \times T^2$ with symmetric embedding in
the $E_8$s and Type IIA (IIB) on Calabi-Yau
spaces $X$ ($X^*$).  
To find simple tests of the duality we shall consider
perturbative CHL compactifications.  Then,
$X$ will be a $K3$ fibration \klm, and the IIA dual
will be a $Z_2$ orbifold of $X$.
Using the adiabatic argument, we expect the $Z_2$ to 
act as in \chldual\ on the $K3$ fibre of the CY manifold.

The relevant CY manifolds on which such $Z_2$ actions are
to be expected are constructed as follows. We consider the
most general elliptic fibre $X_6(3,2,1)$ in affine coordinates

\eqn\fibre{y^2 + x^3 + a_1 x y + a_2 x^2 + a_3 y + a_4 x + a_6=0.}
Here we view the $a_i$ as functions of the coordinate
$t$ of $\IC$. Locally, the Kodaira types $I$ of minimal singularities
of the fibre over $(x,y,t)=(0,0,0)$ were analysed using a
generalized Tate's algorithm in \bikmsv . The type is specified
essentially\foot{Up to some factorisation conditions, which have to be
imposed
in a few cases as extra constraints.} by the degree of $a_i$ in $t$, see
Table 2 of \bikmsv .

To get a compact Calabi Yau model we instead view the $a_i$ as
sections of  
line-bundles over $\IP^1\times \IP^1$ of type
${\cal O}(k_i)\otimes{\cal O}(k_i)$ with $k_i=2,4,6,8,12$ respectively.
If we forget about one of the $\IP^1$s, this yields a $K3$. If
$(t:t')$ are homogeneous coordinates on the remaining $\IP^1$, the
singular
fibres which occur at the north and south pole
are determined by the lowest degrees of the homogenous
polynomials $a_i$ of degree $k_i$ in $t$ and $t'$ respectively
(as in
Table 2 of \bikmsv).

Now, let $(t:t')$ and $(s:s')$ be the homogeneous parameters of the
first and the second $\IP^1$. We denote the global model by
$I_w {I_n\atop I_s} I_e$ were we understand that singular
fibres of type $(I_n,I_s)$ appear at $t=0$ and $t'=0$ with
$(s,s')$ generic and $(I_e,I_w)$ appear $s=0$ and $s'=0$ with $(t,t')$
generic. The Newton polyhedron $\Delta$ of \fibre\ is the convex hull of

\eqn\polyhedra{\eqalign{&(0,0,0,2), \ (0,0,3,0) \cr
&(a^{(w)}_i,a_i^{(n)},v_i),\ (a^{(w)}_i,k_i-a_i^{(s)},v_i),\cr
&(k_i-a^{(e)}_i,a_i^{(n)},v_i),\ (k_i-a^{(e)}_i,k_i-a_i^{(s)},v_i),}}
with $i=1,2,3,4,6$ and $v_i=(1,1),(2,0),(0,1),(1,0),(0,0)$ shifted
by $(-1,-1,-1,-1)$,
 and always contains the origin\foot{For the fibre
$K3$ the corresponding polyhedra are obtained by deleting the
first or second entry of these vectors.}.
Let $\Lambda$ be the coarsest lattice, which contains all integral
points inside $\Delta$, $V=\Lambda_{\IR}$ the real extension,
$\Lambda^*$ and $V^*$ the dual lattice and vector space respectively
\batyrev .
The polyhedra are reflexive if
\eqn\dualpoly{\Delta^*=\{ x \in V^* \|
\langle x,y\rangle\ge -1,\forall y \in \Delta \}}
is a lattice polyhedron in $\Lambda^*$ and in this case $\Delta$
defines a Calabi-Yau or $K3$ space \batyrev. For affine patches
reflexivity implies the condition for having canonical hypersurface
singularities given 
in \reidI,\reidII\ and the toric description gives a
straightforward prescription to resolve them \reidII.
Of course reflexivity is stronger, i.e. it is not always possible
to compactify the affine patches to a $c_1=0$ manifold.

Points in $\Delta^*$ which are not at codimension one correspond
to divisors in $\IP_{\Delta^*}$, which intersect with the hypersurface
and give rise to divisors in the CY. Most commonly in the examples
below the intersection with the hypersurface yields
one irreducible divisor 
on the CY. i.e. there is one $(1,1)$-form for each point in codim
2 and 3 in $\Delta^*$. In general the number of irreducible divisors 
is given by the number of points interior to the dual face plus one, see
\batyrev\ for details. There are $\kappa$ homology relations in
the full set of divisors, where $\kappa={\rm dim}_{\IC}(X)+2$.  
So in order to count the independent $(1,1)$-forms in the CY ($K_3$), 
one must subtract 5(4) from the number of these divisors.

If we write the resolved \fibre~in Batyrev-Cox coordinates
its specialization to any affine patch of the toric variety can
be neatly displayed. In particular it is easy to identify the
exceptional divisors, which are needed to resolve the singularities
of the affine equations \fibre . The form of the polynomial is
\eqn\batcox{p=\sum_{i} b_i \prod_{j} x_j^{\langle
\nu_i,\nu_j^*\rangle+1},}
where the sum runs over the relevant points $\nu_i$ of $\Delta$ and the
product over the relevant points $\nu^*_j$ of $\Delta^*$.
E.g. the affine patch with the \fibre~singularity in one patch
of the $\IP^1$ is obtained by setting all variables in \batcox~to
one except of the ones associated to $\nu^*_x,\nu^*_y,\nu^*_t$ (see
the Appendix). The mirror polynomial has the analogous definition with
$\Delta$ and $\Delta^*$ exchanged.

Points in $\Delta$ correspond to monomials in the defining
equation of the hypersurface \batcox~and their coefficients
correspond generically to independent complex structure deformations.
They in turn are in one to one correspondence with the $(2,1)$ forms
of the CY. But also in $\Delta$ the codimension one points do
not contribute to the complex structure deformations and hence
to $(2,1)$-forms. The reason here is that one can use the
projective invariance group of the toric ambient space
$\IP GW(\Delta^*,\IC)$ to set (gauge) all their coefficients 
to constant values (zero). We shall call points not on 
codimension one relevant points.
As \batcox~is quasihomogeneous, i.e. scale invariant under a overall
rescaling of the $x_i$, we 
shall rather consider\foot{E.g.for the elliptic curve 
$X_6(1,2,3)$, $GW(\Delta^*,\IC)$ is 
generated by the weight compatible transformation 
$t\mapsto a t$, $x\mapsto b_1 x + b_2 t^2$ and 
$y\mapsto c_1 y + c_2 t x+ c_3 t^3$
with complex coefficients. Three parameters correspond to the $(\IC^*)^3$, 
while 
three can be used 
to set the coefficients of the three codim 1 points $k_1,k_2,k_3$ 
(cff. Appendix A) to zero.}  the $GW(\Delta^*,\IC)$,
which contains a $(\IC^*)^\kappa$ action of
independent rescalings of the variables $x_j$. This leaves
\batcox~invariant upon a $(\IC^*)^\kappa$  rescaling of
the coefficients $a_i$. We can (gauge) fix this freedom of
rescaling by setting $\kappa$ additional coefficients $a_i$ to one.
So generically the number of $(2,1)$-forms is the number of relevant
points in $\Delta$ minus $\kappa$. Analogous to the situation 
with $\Delta^*$, certain points in $\Delta$ can correspond to several
$(2,1)$-forms.  
These points correspond
to the deformation modes of curves (of genus $g$) of singular loci, 
with a $\IC^2/Z_r$ action on the normal bundle.  The
resolution of the curves supports 
$g(r-1)$ additional $(2,1)$-forms, which do not correspond to 
deformation parameters in \batcox.  Again the number of additional 
(2,1) forms is given by the number of interior points in the dual 
face cff. \batyrev .
The gauge fixing of the coefficients determines the dimension of 
the moduli space. There remains however generically a 
(huge) discrete subgroup of $GW(\Delta^*,\IC)$ of invariances of 
\batcox, the so called $R$ symmetries. As we shall see, the $Z_2$ 
symmetry on the coordinates lifts part of the $R$ symmetries of 
the moduli space of the invariant sector.

According to \klm, we can identify the complexified volume of e.g.
the $(w,e)$ $\IP^1$ with the heterotic dilaton and as long as we
are interested only in generic perturbative gauge enhancements we
keep the fibre over this $\IP^1$ generic, i.e. $I_w=I_e=I_0$,
which implies $a^{(w)}_i=a^{(e)}_i=0$. Moreover as we want to
have cases with symmetric unhiggsing of the gauge group to perform
the $Z_2$ modding we look at models $X(I):=I_0 {I\atop I} I_0$.
We list in Table 2 various cases.

The $\IZ_2$ symmetry acts by by exchanging
\eqn\sigmaact{ \sigma: t \rightarrow t', \quad t'\rightarrow t,
                              \quad  y\rightarrow -y.}
It is obvious that \polyhedra~has two symmetry planes associated
with the exchange of $(t,t')$ and $(s,s')$. To act with \sigmaact,
we must tune the complex structure parameters $b_i$ in front of
the monomials of $p$, which corresponds to points above and
below the symmetry 
plane, to symmetric or antisymmetric values (depending on whether they
multiply $y$). We must take the normal subgroup $N$ of 
$GW(\Delta^*,\IC)$ w.r.t. to \sigmaact~to subtract the
reparametrisation invariance. If we can globally diagonalize 
\sigmaact~, as e.g. in the example below, it is easy to see 
that the dimension of $N$ is the number of invariant points in 
codimension one. However for general actions,
that is not the case 
and we have to explicitly determine $N$ to count the independent 
$\sigma$-invariant $(2,1)$ forms.

In fact due to \dualpoly~the dual polyhedron also has two such symmetry
planes and we have to enforce symmetric values of the vector moduli
in the mirror polynomial as well. The counting of invariant $(1,1)$-forms
proceeds in a way
similar to the counting in 
the $\Delta$ polyhedron, and is summarized in Table 2.

For all entries except the $E_8^2$ case, one should compare to the
$n=8$ compactification of \S2 (with no small instantons).
For the $E_8^2$ case, all of the instantons are small and one is
in the $N=12$ case of \S2.  Then, $G=E_8$ so there are
$20+12=32$ hypermultiplets coming from the $K3$ moduli and
positions of the (wrapped) fivebranes, and $12+8+3 = 23$ vector
multiplets at generic points in the Coulomb branch.
In all cases, using
\eqn\spect{n_{V} = h^{1,1}_{inv},~~n_{H} = h^{2,1}_{inv} + 1}
we find agreement between the CHL expectations and the
numbers of $Z_2$ invariant cohomology classes.

Table 2: Symmetric $n=8$ models and $Z_2$ invariant sector. 
The numbers in the brackets 
correspond to non-toric deformations associated to curve singularities in $X$.    
\bigskip
\begintable
$ X(I)$      | Group   | $h^{1,1}$  | $h^{1,1}$ | $h^{1,1}_{inv}$ |
$h^{2,1}_{inv}$ \elt
$X(II^*)$    |$(E_8)^2$|  43(22)    |  43 (0)   |    23(0)        |
31(10)          \elt
$X(III^*)$    |$(E_7)^2$|   17(0)   |  61(0)    |    10(0)        |
40(0)           \elt
$X(IV^{*s})$  |$(E_6)^2$|   15(0)   |  75(0)    |    9(0)         |
47(0)           \elt
$X(IV^{*ns})$ |$(F_4)^2$|   11(0)   | 107(20)   |    7(0)         |
53(10)          \elt
$X(I_2^{*ns})$|$(SO(11))^2$|13(0)   | 85(14)    |    8(0)         |
52(7)           \elt
$X(I_1^{*s})$ |$(SO(10)^2$| 13(0)   |  85(0)    |    8(0)         |
52(0)           \elt
$X(I_1^{*ns})$|$(SO(9))^2$|  11(0)  | 107(5)    |    7(0)         |
63(5)           \elt
$X(I_0^{*ns})$|$(SO(7))^2)$|  9(0)  | 121(6)    |    6(0)         |
70(3)           \elt
$X(I_0^{*ns})$|$(G_2)^2$  |  7(0)   | 151(20)   |    5(0)         |
85(10)           \elt
$X(I_5)^s)$   |$(SU(5))^2$ | 11(0)  |  91(0)    |    7(0)         |
55(0)           \elt
$X(I_4^s)$    |$(SU(4))^2$|  9(0)   | 121(0)    |    6(0)         |
70(0)           \elt
$X(I_3^s)$    |$(SU(3))^2$|  7(0)   | 151(0)    |    5(0)         |
85(0)           \elt
$X(I_2)$      |$(SU(2))^2$|  5(0)   | 185(0)    |    4(0)
|102(0)           \elt
$X(I_0)$      | no gen.   |  3(0)   | 243(0)    |    3(0)
|131(0)
\endtable
\bigskip

Let us discuss the $X(I_0)$ case in some detail. From \polyhedra~we see
that
$\Delta$ has six corners at $\nu_1=(-1, -1, -1, 1)$, $\nu_2=(-1, -1, 2,
-1)$,
$\nu_3=(-1, 11, -1, -1)$, $\nu_4=(11, -1, -1, -1)$,
$\nu_5=(11, 11, -1, -1)$ and $\nu_6=(-1, -1, -1, -1)$.  The lattice
$\Lambda$ is spanned by standard unit vectors $e_i$ in $\IR^4$.
We can diagonalize the action of \sigmaact~without changing the
shape of $\Delta$. On the new coordinates $s,s'$, $\tilde t$,
${\tilde t}',z,x,y$, the $Z_2$
acts by $x_i\mapsto \exp(2 \pi i r_i) x_i$ with
$r={1\over 2}(0,0,0,1,0,0,1)$. This $Z_2$ orbifoldisation does
commute now with the action of the algebraic torus defining the toric
ambient space and hence can be taken by considering the quotient
lattice $\hat \Lambda=\Lambda/\IZ$ of $\Lambda$ spanned by
\eqn\latticequot{\hat e_1=e_1, \
\hat e_2=e_2+e_4, \
\hat e_3=e_1, \
\hat e_4=2 e_4}
with ${\hat \Lambda}^*$ the dual to $\hat \Lambda$, as in \batyrev.

The resolved orbifold Calabi-Yau $\hat X={\widehat {X/\IZ_2}}$
is defined by the old polyhedra $(\Delta,\Delta^*)$ in the coarser
lattice
$\hat \Gamma=\Gamma/\IZ_2$ and the finer dual lattice
${\hat \Lambda}^*$. Note that $\Lambda^*={\hat \Lambda}^*/\IZ_2$
and this defines an action of the $\IZ_2$ on ${\hat \Lambda}^*$,
which in turn can be used to define the dual orbifoldization
${\widehat {{\widehat {X/\IZ_2}}/\IZ_2}}=X$.  The invariant
$(2,1)$-forms ($(1,1)$-forms) correspond to those points of
$\Delta$ ($\Delta^*$), which are on the coarser lattices
$\hat \Lambda$ ($\Lambda^*$)\foot{Hence it seems difficult to 
find such group actions which diminish both the numbers of
$(1,1)$ and 
$(2,1)$ forms. However frequently one can deform the (vector) moduli 
space to a point, where a sufficient number of (vector) moduli 
become non-toric, so that their number now indeed depends on the points
in the dual lattice. E.g. if 
we set for the $X(I_{A_1})$ model the perturbations which
corresponds to $\nu^{*}_t$ and $\nu^{*}_{t'}$ to zero,
we get the cohomology
$h^{1,2}=185(0)$ and $h^{1,1}=5(1)$.  Then 
under \latticequot~we get the CHL 
cohomology in the 
invariant sector: $h^{1,2}_{inv}=102(0)$ and $h^{1,1}_{inv}=4(0)$. 
Similarly the $X(I_{E_8})$ example 
is at a point in the moduli space where 
\latticequot~gives the CHL spectrum.}. 

Points in $\hat \Lambda^*$,
but not in $\Lambda^*$ correspond to the twisted $(1,1)$-forms
of the original, and points in $\Lambda$ but not in
$\hat \Lambda$ correspond to the twisted $(2,1)$-forms of the dual
orbifold.

In the particular case of $X(I_0)$, $h^{inv}_{1,1}(\hat X)=3$  and
$h^{inv}_{2,1}(\hat X)=131$ while $h_{1,1}(\hat X)=9$,
$h_{2,1}(\hat X)=153$.  Keeping the invariant modes, we
find agreement with the expectation from the
CHL construction.

The vector moduli space of the CHL string is described by the
deformations of the mirror polynomial. The mirror manifold of $X(I_0)$
can be itself obtained by a quotient of a group $G$ of order 72 on $X$. 
The quotient lattice
$\Lambda_M$ is spanned by $e^M_1=e_1+e_2+e_3$, $e_2^M=12 e_2$, $e_3^M=3
e_3$
and $e^M_4= 2 e_4$. The mirror polynomial is defined by \batcox~with $i$
running
over the relevant points of $\Delta$, which are in $\Lambda^M$ and $j$
running over
the relevant points in $\Delta^*$
\eqn\mirrorpoly{p=x_0(a_1 y^2+a_2 x^3 + z^6\{a_3 (s s' t t')^6+ a_4 (s
t)^{12} +
a_5 (s t')^{12}+ a_6 (s' t)^{12}+ a_7 (s' t')^{12}\}+ a_0 x y s s' t
t'),}
where the coordinates $(s,s',t,t',z,x,y)$ for the CHL mirror 
are identified by the action of $G$, which is generated by 
$r_1={1\over 12}(-1,0,0,1,0,0,0)$, 
$r_2={1\over 3}(0,0,0,1,-1,0,0)$ but not by 
$r_3={1\over 2}(0,0,0,1,0,0,1)$ as it would be for the mirror 
of the $X(I_0)$ model. As a consequence the CHL moduli space 
is a double covering of the one of the $X(I_0)$ model.

Another interesting example \kv, which is not in Table 1,
comes from the $E_8 \times E_8$
heterotic theory on $K3\times T^2$ with symmetric $SU(2)$
instanton embedding $(n_1,n_2)=(10,10)$ in the $E_8\times E_8$ and
$n=4$ in the \lq\lq stringy" $SU(2)$ of the $T^2$, which
we take to be at an enhanced symmetry point.
It has $2(20-3)$ hypermultiplets from the instantons in the
$E_8$s, $8-3$ from the ones in the $SU(2)$, $20$ from the gravitational
sector and $2(3\cdot 56-133)$ from higgsing the $E_7$.
Orbifolding by the CHL $Z_2$, we find that
the hypermultiplet
counting for the CHL string should be $17+5+20+35$, while there should
be $2$ vector multiplets. The polyhedron for the CHL dual
is spanned by
$\nu_1=(11,-1,-1,-1)$,
$\nu_2=(-1,5,-1,-1)$,
$\nu_3=(-1,-1,5,-1)$,
$\nu_4=(-1,-1,-1,1)$, and
$\nu_5=(-1,-1,-1,-1)$. After the quotient by
\latticequot~we get $h^{inv}_{1,1}(\hat X)=2$  and
$h^{inv}_{2,1}(\hat X)=76$ and the resolved cohomology is
$h_{1,1}(\hat X)=5$ and
$h_{2,1}(\hat X)=101$. Again, the invariant cohomology is
in accord with the expectation for the CHL spectrum.

\bigskip
\centerline{\bf{Acknowledgements}}

We would like to thank
O. Aharony, J. Schwarz, and E. Silverstein for helpful 
discussions on related subjects.
The research of S.K. and Y.O. was supported in part by
NSF grant PHY-95-14797 and by the DOE under contract
number DE-AC03-76SF00098.  S.K. is also supported by a
1997 DOE OJI Award.

\newsec{Appendix}

Although \polyhedra~and \dualpoly~define $\Delta^*$
for all cases,
here we give a more concrete description in a convenient
basis (see also \bikmsv\cf)\foot{In this basis one can easily visualize
the
$K3$ polyhedron,
see \cf. This basis is related to the one which comes out of direct
application of
\dualpoly~by the matrix $\left(\matrix{1&0&0&0\cr 0&1&0&0\cr 2&0&1&0\cr
3&0&0&1}\right)$}
Let $k_y=(0,1)$,
$k_x=(1,0)$,
$k_0=(0,0)$,
$k_1=(0,-1)$,
$k_2=(-1,-1)$,
$k_3=(-1,-2)$,
$k_z=(-2,-3)$
be the Newton polyhedron of the $X_6(1,2,3)$
elliptic curve and $\nu^{n}_{k_i}=(0,n,k_i)$. Then $\Delta^*$
always involves the
relevant points $\nu_0=(0,0,0,0)$,
$\nu^*_s=(1,0,-2,-3)$, $\nu^*_{s'}=(-1,0,-2,-3)$,
$\nu^*_{t}=(0,1,-2,-3)$,
$\nu^*_{t'}=(0,-1,-2,-3)$, $\nu^*_{z}=(0,0,-2,-3)$,
$\nu^*_{x}=(0,0,1,0)$,
$\nu^*_{y}=(0,0,0,1)$, which describe the  dual polyhedron and hence
the vector moduli space of $X(I_0)$. The unhiggsing of $\Delta^*$
adds the following points.

Table 3: Dual Polyhedra for the symmetric cases
\bigskip
\begintable
$ X(I)$      | group  |additional points   \elt
$X(II^*)$    | $(E_8)^2$|
$\nu^{\pm 6}_{k_z} \ldots \nu^{\pm 2}_{k_z}, \nu^{\pm 4}_{k_3},
\nu^{\pm 3}_{k_2}, \nu^{\pm 2}_{k_1}, \nu^{\pm 1}_{k_0}
$\elt
$X(III^*)$    |$(E_7)^2$| $\nu^{\pm 4}_{k_z} \ldots \nu^{\pm 2}_{k_z},
\nu^{\pm 3}_{k_3}, \nu^{\pm 2}_{k_2}, \nu^{\pm 2}_{k_1}, \nu^{\pm
1}_{k_0}$,
     \elt
$X(IV^{*s})$  |$(E_6)^2$| $\nu^{\pm 3}_{k_z},\nu^{\pm 2}_{k_z},
\nu^{\pm 2}_{k_3}, \nu^{\pm 2}_{k_2}, \nu^{\pm 1}_{k_1},
\nu^{\pm 1}_{k_0}$,     \elt
$X(IV^{*ns})$ |$(F_4)^2$| $\nu^{\pm 3}_{k_z},\nu^{\pm 2}_{k_z},
\nu^{\pm 2}_{k_3}, \nu^{\pm 1}_{k_1},$     \elt
$X(I_2^{*ns})$|$(SO(11))^2$| $\nu^{\pm 2}_{k_z},\nu^{\pm 2}_{k_3},
\nu^{\pm 1}_{k_2}, \nu^{\pm 2}_{k_1},$
  \elt
$X(I_1^{*s})$ |$(SO(10)^2$|$\nu^{\pm 2}_{k_z},\nu^{\pm 2}_{k_3},
\nu^{\pm 1}_{k_2}, \nu^{\pm 1}_{k_1},\nu^{\pm 1}_{k_0} $
    \elt
$X(I_1^{*ns})$|$(SO(9))^2$|  $\nu^{\pm 2}_{k_z},
\nu^{\pm 2}_{k_3},\nu^{\pm 1}_{k_2}, \nu^{\pm 1}_{k_1},$
  \elt
$X(I_0^{*ns})$|$(SO(7))^2)$|$\nu^{\pm 2}_{k_z},
\nu^{\pm 1}_{k_2},\nu^{\pm 1}_{k_1}, \nu^{\pm 1}_{k_1},$
   \elt
$X(I_0^{*ns})$|$(G_2)^2$  | $\nu^{\pm 2}_{k_z},\nu^{\pm 1}_{k_1},$
   \elt
$X(I_5)^s)$   |$(SU(5))^2$ |$\nu^{\pm 1}_{k_3},
\nu^{\pm 1}_{k_2},\nu^{\pm 1}_{k_1},\nu^{\pm}_{k_0},$   \elt
$X(I_4^s)$    |$(SU(4))^2$|
 $\nu^{\pm 1}_{k_3},\nu^{\pm 1}_{k_2},\nu^{\pm 1}_{k_1}$
   \elt
$X(I_3^s)$    |$(SU(3))^2$|
$\nu^{\pm 1}_{k_3},\nu^{\pm 1}_{k_2}$
\elt
$X(I_2)$      |$(SU(2))^2$|
$\nu^{\pm 1}_{k_3}$
\elt
$X(I_0)$      | no gen.   |  -
\endtable
\listrefs

\end